\documentclass[12pt,eadjoint tfnpsf]{article}

\catcode`\@=11
\@addtoreset{equation}{section}

\global\arraycolsep=1pt

\usepackage{amsbsy,amssymb,latexsym,amsfonts, amsmath}
\usepackage{mathrsfs}
\usepackage{graphicx}

\RequirePackage[dvips,usenames]{color}
\definecolor{fireblick}{rgb}{0.698039,0.133333,0.133333}

\newcommand{\beq}{\begin{equation}}
\newcommand{\eeq}{\end{equation}}
\newcommand{\bea}{\begin{eqnarray}}
\newcommand{\eea}{\end{eqnarray}}

\newcommand{\CB}{{\mathcal B}}

\newcommand{\CE}{{\mathcal E}}

\newcommand{\CN}{{\mathcal N}}
\newcommand{\CO}{{\mathcal O}}

\newcommand{\CZ}{{\mathcal Z}}

\newcommand{\M}{{\mathfrak M}}

\newcommand\C{{\mathbb{C}}}

\renewcommand{\thefootnote}{\fnsymbol{footnote}}

\begin{document}
%
%
\begin{titlepage}

\begin{flushright}
\normalsize
~~~~
SISSA  39/2011/EP-FM\\
\end{flushright}

\vspace{60pt}

\begin{center}
{\huge Gauge Theories on ALE Space }\\ 
\vspace{16pt}
{\huge  and }\\ 
\vspace{16pt}
{\huge Super Liouville Correlation Functions}\\
\end{center}

\vspace{25pt}

\begin{center}
{
Giulio Bonelli${}^{\heartsuit, \clubsuit}$, Kazunobu Maruyoshi${}^{\heartsuit}$ 
and Alessandro Tanzini${}^{\heartsuit}$
}\\
%
\vspace{20pt}
%
${}^\heartsuit$ 
{\it International School of Advanced Studies (SISSA) \\via Bonomea 265, 34136 Trieste, Italy 
and INFN, Sezione di Trieste }\\
\vspace{7pt}
${}^\clubsuit$ 
{\it I.C.T.P. -- Strada Costiera 11, 34014 Trieste, Italy}\\
\end{center}
%
\vspace{17pt}
\begin{center}
Abstract\\
\end{center}
We present a relation between $\mathcal{N}=2$ quiver gauge theories on the ALE space
$\mathcal{O}_{\mathbb{P}^1}(-2)$ and correlators of $\mathcal{N}=1$ super Liouville conformal
field theory, providing checks in the case of punctured spheres and tori.
We derive a blow-up formula for the full Nekrasov partition function and show
that, up to a $U(1)$ factor, the $\mathcal{N}=2^*$ instanton partition function is given by the product
of the character of $\widehat{SU}(2)_2$ times the super Virasoro conformal block on the torus with one puncture. 
Moreover, we match the perturbative gauge theory contribution with super Liouville three-point functions.  
%
\vspace{20pt}\\
\footnotesize{Mathematical Subject Classification (2010): 81T40, 81T60.}\\
\footnotesize{Keywords: conformal field theory, supersymmetric gauge theory.}

\setcounter{footnote}{0}
\renewcommand{\thefootnote}{\arabic{footnote}}

\end{titlepage}

\section{Introduction}
\label{sec:intro}
  A relation between $\mathcal{N}=2$ superconformal quiver gauge theories 
  on $\mathbb{C}^2$ and Liouville conformal field theory correlators was pointed out in \cite{AGT} 
  building on M-theory geometrization of non-perturbative dualities \cite{N=2}.
  In this paper we show that an analogous relation holds between gauge theories on an ALE space 
  and $\mathcal{N}=1$ super Liouville conformal field theory. 
  In particular we analyze $\mathcal{N}=2$ gauge theories on the cotangent bundle 
  of the two-sphere $\mathcal{O}_{\mathbb{P}^1}(-2)$, 
  namely the minimal resolution of the $\mathbb{C}^2/\mathbb{Z}_2$ orbifold, 
  and show that its perturbative and instanton sectors provide 
  respectively the three-point functions and conformal blocks of $\mathcal{N}=1$ super Liouville. 
  This latter relation concerning the conformal blocks of super Virasoro algebra 
  was recently discussed in \cite{BF,BMT,BBB} in some particular cases. 
  
  An M-theory perspective on such a correspondence was elaborated in \cite{NT} by suggesting that $N$ M5-branes 
  on $\mathbb{C}^2/\mathbb{Z}_m$ should give rise to a two-dimensional system 
  with $U(1)$, $\widehat{SU}(m)_N$ and $m$-th para-$W_N$ symmetries.
  This proposal was checked by computing the central charge 
  arising from the anomaly polynomial of the multiple M5-brane system following the approach of \cite{BT,ABT}. 
  
  We present a detailed analysis of $\mathcal{N}=2$ quiver gauge theories 
  and find a blow-up equation expressing the full partition function on the ALE space
  in terms of $\C^2$ partition functions. We then compare with super Liouville correlators on punctured spheres and tori.
  A byproduct of our analysis is that the partition function of $\mathcal{N}=2^*$ theory, 
  corresponding to the torus with one puncture, 
  reproduces the character of $\widehat{SU}(2)_2$ times the super Virasoro conformal block
  and also the contribution from $U(1)$, confirming the general arguments of \cite{NT}.
  
  In Section \ref{sec:gauge} we discuss the details of the full Nekrasov function on $\mathcal{O}_{\mathbb{P}^1}(-2)$ 
  for quiver gauge theories.
  In Section \ref{sec:superLiouville} and \ref{sec:superLiouville5} 
  we work out the relation with the four and five points super Liouville correlators 
  on the sphere respectively.
  In Section \ref{sec:superLiouvilletorus} we study the same relation for the torus with one puncture 
  and highlight the relation with the character of $\widehat{SU}(2)_2$.
  Finally in Section \ref{sec:conclusion} we present our conclusions and discussions on further directions, 
  and collect some useful identities in the Appendices.

\section{Gauge theories on ALE space}
\label{sec:gauge}
  In this section we consider gauge theories on the $\CO_{\mathbb{P}^1}(-2)$ complex surface, which is
  the minimal resolution of the $A_{1}$ singularity $\C^2/\mathbb{Z}_2$.
  In subsection \ref{subsec:instanton} and \ref{subsec:pert}, 
  we will consider the instanton and perturbative parts of the partition function respectively.

\subsection{Instanton partition function}
\label{subsec:instanton}
  
  The compactification of the moduli space $\widetilde\M(r,k,n)$ of rank $r$ instantons on $\CO_{\mathbb{P}^1}(-2)$ 
  can be described in terms of the moduli space of framed torsion free sheaves $(\CE,\Phi)$, 
  where $\Phi$ is the framing on a suitable divisor, on the global quotient $\mathbb{P}^2/\mathbb{Z}_2$
  with minimal resolution of the singularity at the origin \cite{nakamoscow}.
  The resulting variety corresponds to a ``stacky" compactification of ${\mathcal O}_{\mathbb{P}^1}(-2)$ 
  obtained by adding a divisor $\tilde C_\infty\simeq \mathbb{P}^1/\mathbb{Z}_2$ 
  over which the framing is defined \cite{BPT}.
  This variety is an algebraic Deligne-Mumford stack $X_2$ 
  whose coarse space is the second Hirzebruch surface $\mathbb{F}_2$.
  The crucial point that lead to consider this compactification is that on $X_2$ 
  one has line bundles with half-integer first Chern class supported on the exceptional divisor. 
  From the physics viewpoint, this allows to include the contribution of anti-self-dual gauge connections 
  with nontrivial holonomy at infinity, see \cite{BPT} for details.
  The moduli space $\widetilde\M(r,k,n)$ is characterized by the rank $r$, 
  the first Chern class of the $\CE$ sheaf $c_1(\mathcal{E})= k C$, 
  where $C$ is the exceptional divisor resolving the singularity at the origin, and the discriminant
    \bea
    n
     =     \int \left( c_2(\CE)-\frac{r-1}{2r}c_1^2(\CE) \right) \ .
    \eea
  Since the exceptional divisor squares to $-2$, we have $\int c_1(\CE) \wedge c_1(\CE) = - 2k^2$. 
  Then the instanton action, given by the integral of the second Chern character, reads
    \bea
    S_{inst}
     =     \int \left( c_2(\CE) - \frac{1}{2} c_1(\CE)^2 \right)
     =     n + \frac{k^2}{r}.
           \label{ch2}
    \eea
  We underline that $k$ is in general half integer due to the ``stacky'' compactification of the ALE space \cite{BPT}.
  Indeed, half-integer classes take into account anti-self-dual connections which asymptote flat connections 
  with nontrivial holonomy at infinity. 
  These will play an important r\^ole in the correspondence with super Liouville theory, 
  being related to the Neveu-Schwarz (NS) and Ramond (R) sectors respectively. 
  In the following we will concentrate on $k \in \mathbb{Z}$ which corresponds to the NS sector.

  The evaluation of the instanton partition function can be obtained by using the localization techniques 
  developed in \cite{Nekra,NO,Flupo,BFMT,NY-I}.    
  The torus action $T$ on the instanton moduli space is given by the Cartan gauge rotations parametrized 
  in terms of the vevs of the vector multiplet scalars $a_\alpha$ ($\alpha = 1, \ldots, r$) 
  and the space rotations with angles $\epsilon_{1,2}$.
  This acts on ${\mathcal O}_{C}(-2)$ as $T: [z:w]\to [t_1z:t_2w]$ on the exceptional divisor $C$ 
  and as $(z_1,z_2)\to (t_1^2 z_1, t_2^2 z_2)$ on the fibers over it, 
  where $t_{1,2} = e^{\epsilon_{1,2}}$. 
  On the exceptional divisor the fixed points of the torus action are given by $w=0$ and $z=0$.
  
  We are interested in the fixed points of the instanton moduli space under the above torus action. 
  As discussed in \cite{Sasaki,BPT},
  these are given by ideal sheaves $I_\alpha$ twisted by $\mathcal{O}(k_\alpha C)$ line bundles, $\alpha=1,\ldots,r$, 
  and are specified in terms of $\vec{k} = (k_1, \ldots, k_r)$ and a pair of Young diagrams $\vec{Y}_1, \vec{Y}_2$,
  where $\vec{Y}_1$ (resp. $\vec{Y}_2$) parametrizes the contribution from the fixed point at $w=0$ (resp. $z=0$).
  In the following we will use the compact notation $I_\alpha\left(  k_\alpha C \right)$.
  The fixed points data have to satisfy the following relations
    \bea
    n
     =     |\vec{Y}_1| + |\vec{Y}_2| + \frac{1}{r} \sum_{\alpha < \beta}(k_{\alpha \beta})^2,
           ~~~~~
    k
     =     \sum_\alpha k_\alpha,
           \label{relationnk}
    \eea
  where $k_{\alpha \beta} = k_\alpha - k_\beta$.
  
  According to the decomposition $\CE = \oplus_\alpha I_\alpha(k_\alpha C)$, the tangent space at the fixed points 
  $T_{(\CE,\Phi)} \widetilde\M(r,k,n) = {\rm Ext}^1\left(\CE,\CE(-\tilde C_\infty)\right)$ is decomposed as
  $\oplus_{\alpha,\beta}{\rm Ext}^1 \left(I_\alpha (k_\alpha C),I_\beta(k_\beta C-\tilde C_\infty)\right)$. 
  It can be shown (see \cite{BPT} for details) that the non vanishing contributions to the above are given by
  ${\rm Ext}^1 \left({\cal O}(k_\alpha C),{\cal O}(k_\beta C-\tilde C_\infty)\right)$, 
  $t_1^{2k_{\alpha\beta}}{\rm Ext}^1 \left(I_\alpha^1,I_\beta^1\right)$ and
  $t_2^{2k_{\alpha\beta}}{\rm Ext}^1 \left(I_\alpha^2,I_\beta^2\right)$.
  
  The corresponding T-module structure of the tangent space at the fixed points, 
  corresponding to the vector multiplet contribution, is the following 
    \bea
    \label{tangent}
    \chi^{{\rm vector}}(\vec{a})
    &=&    \sum_{\alpha ,\beta=1}^r \left(L_{\alpha ,\beta}(t_1,t_2)
         + t_1^{2k_{\alpha \beta}} N_{\alpha ,\beta}^{\vec{Y}_1} (t_1^2, t_2/t_1)
         + t_2^{2k_{\alpha \beta}}
           N_{\alpha ,\beta}^{\vec{Y}_2}(t_1/t_2,t_2^2)\right),\label{tangent_space}
    \eea 
  where $L_{\alpha ,\beta}(t_1,t_2)$ is given by 
    \begin{equation}
    L_{\alpha, \beta} (t_1,t_2) 
     =     e_\beta\, e_\alpha ^{-1} \times \left\{ \begin{array}{ll}
           \displaystyle{\sum_{{\scriptstyle i,j\ge 0, \ i+j \le 2(k_{\alpha\beta}-1)} 
           \atop {\scriptstyle i+j-2k_{\alpha\beta} \equiv 0 \ {\rm mod}~ 2}}
           t_1^{-i} t_2^{-j} } , &\quad k_{\alpha \beta} > 0   \\
           \displaystyle{\sum_{{\scriptstyle i,j\ge 0, \ i+j \le - 2k_{\alpha\beta}- 2}
           \atop {\scriptstyle i+j+2+2k_{\alpha\beta}\equiv 0 \ {\rm mod}~ 2}}
           t_1^{i+1} t_2^{j+1} }, &\quad k_{\alpha \beta} < 0   \\
           0 , &\quad  k_{\alpha \beta} = 0
           \end{array} \right.
           \label{Lmag0}
    \end{equation}
  and
    \bea
    N_{\alpha ,\beta}^{\vec{Y}}(t_1,t_2)
     =     e_\beta e_\alpha ^{-1}\times \left\{\sum_{s \in Y_\alpha }
           \left(t_1^{-l_{Y_\beta}(s)}t_2^{1+a_{Y_\alpha }(s)}\right)+\sum_{s \in Y_\beta}
           \left(t_1^{1+l_{Y_\alpha }(s)}t_2^{-a_{Y_\beta}(s)}\right)\right\}\,,
           \label{R4_tangent_space} 
    \eea
  which is the character of the tangent space corresponding to the vector multiplet on $\mathbb{C}^2$.
  The three terms in (\ref{tangent}) correspond to the three nonvanishing components of the ${\rm Ext}^1$ respectively.  
  In eq.~(\ref{R4_tangent_space}) and in the rest of the paper we use the standard notation $a_Y(s)$ for the 
  (relative) arm and $l_Y(s)$ for the (relative) leg lengths. 
  Note that we choose a convention $e_\alpha = e^{-a_\alpha}$ such that the above character agrees with the 
  usual one in the literature \cite{BFMT, AGT}.
  
  Let us turn to the contribution of bifundamental hypermultiplets 
  in the ($r, \bar{\tilde{r}}$) representation of $U(r) \times U(\tilde{r})$.
  The fixed points are specified in terms of the arrays $\vec{k}$ and $\vec{\tilde{k}}$ 
  and two sets of Young diagrams $\vec{Y}_{1,2}$ and $\vec{W}_{1,2}$.
  The contribution of one massive bifundamental hypermultiplet of mass $m$ can be obtained 
  by generalizing the above procedure and is given by
    \bea
    \label{tangentadjoint}
    \chi^{{\rm bifund}}(\vec{a}, \vec{\tilde{a}}; m)
    &=&  - \sum_{\alpha=1}^r \sum_{\beta'=1}^{\tilde{r}} \left(L_{\alpha,\beta'}(t_1,t_2)
         + t_1^{2k_{\alpha \beta'}} N_{\alpha, \beta'}^{\vec{Y}_1, \vec{W}_1}(t_1^2,t_2/t_1)
         + t_2^{2k_{\alpha \beta'}} N_{\alpha, \beta'}^{\vec{Y}_2, \vec{W}_2}(t_1/t_2,t_2^2) \right) e^{-m},
           \nonumber \\
    & &    
    \eea
  where  
  $a_\alpha$ and $\tilde{a}_{\beta'}$ are the vevs of the vector multiplet scalars of the two gauge groups
  and $L_{\alpha, \beta'}$ can be obtained from (\ref{Lmag0}) by replacing 
  $e_{\beta} e^{-1}_\alpha$ and $k_{\alpha \beta}$ with $e_{\beta'} e^{-1}_\alpha$ and
  $k_{\alpha \beta'} \equiv k_\alpha - \tilde{k}_{\beta'}$ respectively, 
  with $e_{\beta'} = e^{\tilde{a}_{\beta'}}$. Moreover,
  $N_{\alpha, \beta'}^{\vec{Y}, \vec{W}}$ is given by
    \bea
    N_{\alpha, \beta'}^{\vec{Y}, \vec{W}}(t_1,t_2)
     =     e_{\beta'} e_\alpha ^{-1} \times \left\{\sum_{s \in Y_\alpha }
           \left(t_1^{-l_{W_{\beta'}}(s)}t_2^{1+a_{Y_\alpha }(s)}\right)
         + \sum_{s \in W_{\beta'}}
           \left(t_1^{1+l_{Y_\alpha }(s)}t_2^{-a_{W_{\beta'}}(s)}\right) \right\}.
    \eea
 The contribution of adjoint hypermultiplets can be obtained by setting
 $\vec{\tilde{a}} = \vec{a}$ and $\vec{W} = \vec{Y}$. 
 This extends the results of \cite{GL} to the ``stacky'' compactification according to the rules stated in \cite{BPT}.
 The character of the fundamental hypermultiplet of mass $m$ is given 
 by an analogous extensions of the formula in \cite{GL} obtained by setting $\vec{W}=\emptyset$, 
 $\tilde k_{\beta'}=0$ and $\tilde a_{\beta'}=0$ in (\ref{tangentadjoint})
    \bea
    \chi^{{\rm fund}} (\vec{a}, m)
    &=&  - \sum_{\alpha = 1}^r \left(L_{\alpha}(t_1,t_2)
         + t_1^{2 k_\alpha} N_{\alpha}^{\vec{Y}_1}(t_1^2,t_2/t_1)
         + t_2^{2 k_\alpha} N_{\alpha}^{\vec{Y}_2}(t_1/t_2,t_2^2) \right) e^{m - \epsilon_+},
           \label{tangentfund}
    \eea
  where $\epsilon_+ = \epsilon_1 + \epsilon_2$,
    \begin{equation}
    L_{\alpha} (t_1,t_2) 
     =     e_\alpha \times \left\{ \begin{array}{ll}
           \displaystyle{\sum_{{\scriptstyle i,j\ge 0,\, \ i+j \le 2 (k_\alpha - 1)} 
           \atop {\scriptstyle i+j+2 - 2 k_\alpha \equiv 0 \ {\rm mod}~ 2}}
           t_1^{i+1} t_2^{j+1}} , &\quad k_\alpha> 0   \\
           \displaystyle{\sum_{{\scriptstyle i,j\ge 0, \ i+j \le - 2 (k_{\alpha}+1)}
           \atop {\scriptstyle i+j + 2 k_\alpha \equiv 0 \ {\rm mod}~ 2}}
           t_1^{-i} t_2^{-j}}, &\quad k_\alpha < 0   \\
           0 , &\quad  k_{\alpha} = 0
           \end{array} \right.
    \end{equation}
 and\footnote{We note that this formula coincides with the one of $\mathbb{C}^2$,
 up to an overall sign due to a different notation w.r.t. \cite{AGT}.}
    \bea
    N_{\alpha}^{\vec{Y}}(t_1,t_2)
     =     e_\alpha \sum_{s \in Y_\alpha} t_1^{-(i(s)-1)} t_2^{-(j(s) - 1)}
    \eea
 Finally, the contribution of anti-fundamental hypermultiplets is given by
    \bea
    \chi^{{\rm anti-fund}} (\vec{a}, m)
     =     \chi^{{\rm fund}} (\vec{a}, \epsilon_+ - m).
           \label{tangentantifund}
    \eea
  The above discussion can be easily generalized 
  to quiver gauge theories. In this case,
  the fixed points are  described in terms of vectors $\vec{k}^s$ 
  and pairs of Young diagrams $\vec{Y}_1^s, \vec{Y}_2^s$ where $s = 1, \ldots, n$ labels the nodes of the quiver.
  Let $r_s$ and $\vec{a^s}$ be the rank and the Cartan parameters of the $s$-th gauge group.  
  The relations (\ref{relationnk}) have to be satisfied for each $\vec{k}^s$ and $\vec{Y}_1^s, \vec{Y}_2^s$.
  The contributions of vector and matter multiplets for each node are given by the same formulae
  above, written in terms of the $s$ fixed points.   

  By using the above results one can readily compute the instanton part of the Nekrasov partition function 
  of quiver gauge theories on $\mathcal{O}_{\mathbb{P}^1}(-2)$. 
  We resum the contributions by weighting them in terms of 
  the instanton topological action $q^{S_{inst}}=q^{n + \frac{k^2}{2}}$. 
  Then we consider the operator insertion 
  ${\mathcal Z}_{inst}^{ALE} \equiv \left< e^{- \sum_{s=1}^n v_s \int c_1(\CE_s) \wedge c_1(C)} \right>$
  where $C$ is the exceptional divisor and $z_s \equiv e^{2v_s}$
    \bea
    {\mathcal Z}_{inst}^{ALE}(\epsilon_1,\epsilon_2,\vec{a}; q_s, z_s)
     =     \sum_{k^s \in \mathbb{Z}} \left( \prod_{s=1}^n q_s^{\frac{(k^s)^2}{r_s}} (z_s)^{k} \right) 
           {\mathcal Z}_{inst}^{\{k^s\}} (\epsilon_1,\epsilon_2,\vec{a}^s, m_i; q_s),
           \label{zale}
    \eea
  where
    \bea
    {\mathcal Z}_{inst}^{\{k^s\}} (\epsilon_1,\epsilon_2,\vec{a}^s, m_i; q_s)
    &=&    \sum_{\alpha<\beta,{\scriptstyle{\{ k_\alpha^s \}}} \atop {\scriptstyle \sum k_\alpha^s = k^s}}
           \left( \prod_{s=1}^n q_s^{ \frac{(k_{\alpha \beta}^s)^2}{r_s}} \right)
           \ell(\epsilon_1,\epsilon_2, \vec{a}^s, \vec{k}^s, m_i)
           \label{zale2} \\
    & &    ~~~\times
           Z_{inst}^{\mathbb{C}^2} \left(2\epsilon_1,\epsilon_2-\epsilon_1, \vec{a}^s - 2\epsilon_1 \vec{k}^s, q_s \right)
           Z_{inst}^{\mathbb{C}^2} \left(\epsilon_1-\epsilon_2,2\epsilon_2, \vec{a}^s - 2\epsilon_2 \vec{k}^s, q_s \right).
           \nonumber
    \eea
  Here $Z_{inst}^{\mathbb{C}^2} \left(\epsilon_1,\epsilon_2,\vec{a}, q \right)$ 
  is the instanton part of the Nekrasov partition function of the same quiver gauge theory on $\mathbb{C}^2$.
  
  The factor $\ell$ in (\ref{zale2}), depending on the matter content of the gauge theory, 
  is obtained from the characters 
  (\ref{tangent}), (\ref{tangentadjoint}), (\ref{tangentfund}) and (\ref{tangentantifund}).
  Let us spell it out in detail. First of all, let us define
    \bea
    \ell_{\alpha \beta'}(x, k_{\alpha};\tilde{x}, \tilde{k}_{\beta'};m)
     =     \left\{ \begin{array}{ll}
           \displaystyle{\prod_{{\scriptstyle i,j\ge 0, \ i+j \le 2(k_{\alpha \beta'} - 1)} 
           \atop {\scriptstyle i+j-2k_{\alpha \beta'} \equiv 0 \ {\rm mod} ~2}}
           (x - \tilde{x} - i \epsilon_1 - j \epsilon_2 - m) } , &\quad k_{\alpha \beta'} > 0   \\
           \displaystyle{ \prod_{{\scriptstyle i,j\ge 0, \ i+j \le - 2(k_{\alpha \beta'}+1)} 
           \atop {\scriptstyle i+j-2k_{\alpha \beta'} \equiv 0 \ {\rm mod} ~2}}
           (x - \tilde{x} + (i+1) \epsilon_1 + (j+1) \epsilon_2 - m)}, &\quad k_{\alpha \beta'} < 0   \\
           1 , &\quad  k_{\alpha \beta'} = 0
           \end{array} \right.
           \nonumber \\
    & &    \label{ellab}
    \eea
  Similarly, we also define
    \bea
    \ell_{\alpha}(x, k_\alpha)
     =     \left\{ \begin{array}{ll}
           \displaystyle{\prod_{{\scriptstyle i,j\ge 0,\, \ i+j \le 2 (k_\alpha - 1)} 
           \atop {\scriptstyle i+j+ 2 k_\alpha \equiv 0 \ {\rm mod}~ 2}}
           (- x + (i+1) \epsilon_1 + (j+1) \epsilon_2 )} , &\quad k_{\alpha} > 0  \\
           \displaystyle{\prod_{{\scriptstyle i,j\ge 0, \ i+j \le - 2 (k_{\alpha}+1)}
           \atop {\scriptstyle i+j + 2 k_\alpha \equiv 0 \ {\rm mod}~ 2}}
           (- x - i \epsilon_1 - j \epsilon_2)}, &\quad k_{\alpha} < 0   \\
           1 , &\quad  k_{\alpha} = 0
           \end{array} \right.
           \label{ella}
    \eea
  By using these definitions, the $\ell$ factors for the bifundamental and (anti-)fundamental hypermultiplets
  are given by
    \bea
    \ell_{{\rm bifund}}(\vec{a}, \vec{k}; \vec{\tilde{a}}, \vec{\tilde{k}}; m)
    &=&    \prod_{\alpha = 1}^r \prod_{\beta' = 1}^{\tilde{r}} 
           \ell_{\alpha \beta'} (a_{\alpha}, k_{\alpha}; \tilde{a}_{\beta'}, \tilde{k}_{\beta'}; m),
           \\
    \ell_{{\rm fund}}(\vec{a}; m)
    &=&    \prod_{\alpha} \ell_{\alpha} (a_{\alpha} + \epsilon_+ - m, k_\alpha),
           ~~~~
    \ell_{{\rm anti-fund}} (\vec{a}; m)
     =     \prod_{\alpha} \ell_{\alpha} (a_{\alpha} + m, k_\alpha),
           \nonumber
    \eea
  while the adjoint and the vector contributions are
    \bea
    \ell_{{\rm adj}}(\vec{a}; m)
     =     \ell_{{\rm bifund}}(\vec{a}, \vec{k}; \vec{a}, \vec{k}; m), 
           ~~~~~
    \ell_{{\rm vector}} (\vec{a})
     =     \ell_{{\rm adj}} (\vec{a}; 0)^{-1}.
    \eea
  
  Analogously to what discussed in \cite{AGT} for the flat space case, the bifundamental contribution satisfies
    \bea
    \ell_{{\rm bifund}}(\vec{a}, \vec{k}; \vec{\tilde{a}}, \vec{\tilde{k}}; m)
     =     \ell_{{\rm bifund}}(\vec{\tilde{a}}, \vec{\tilde{k}}; \vec{a}, \vec{k}; \epsilon_+ - m),
    \eea
  corresponding to the exchange of the two gauge factors.
  Moreover, the (anti-)fundamental contribution is obtained from the bifundamental one as
    \bea
    \ell_{{\rm bifund}}(\vec{a}, \vec{k}; \vec{\mu}, 0; m)
    &=&    \prod_{f=1}^n \ell_{{\rm fund}} (\vec{a}; m + \mu_f),
           \nonumber \\
    \ell_{{\rm bifund}}(\vec{\mu}, 0; \vec{a}, \vec{k}; m)
    &=&    \prod_{f=1}^n \ell_{{\rm anti-fund}} (\vec{a}; m - \mu_f),
    \eea
  where $\vec{\mu} = (\mu_1, \ldots, \mu_n)$.

\subsection{Classical and perturbative parts of the partition function}
\label{subsec:pert}
  In this subsection, we consider the perturbative part of the partition function.
  As we will show in the following, this is directly related to the three-point functions of super Liouville theory.
  The inclusion of the perturbative contribution will allow us to derive a blow-up formula 
  for the full partition function on the ALE space. 
  Although our result are valid for generic rank, 
  we will focus in this section on the $U(2)$ gauge theory, in which we fix $k=k_1+k_2=0$ and 
  decouple the abelian Coulomb parameter. 
  This is the simplest case for a comparison with super Liouville theory. 
  
  The perturbative part for the vector and the adjoint hypermultiplet contributions of $SU(2)$ gauge theory is
    \bea
    Z^{{\rm vector}}_{pert} (a)
    &=&    \exp \left[- \gamma_{2\epsilon_1, \epsilon_2-\epsilon_1}(2a)
         - \gamma_{2\epsilon_1, \epsilon_2-\epsilon_1}(2a + \epsilon_+)
         - (\epsilon_1 \leftrightarrow \epsilon_2) \right],
           \nonumber \\
    Z^{{\rm adj}}_{pert} (a; m)
    &=&    \exp \left[\gamma_{2\epsilon_1, \epsilon_2-\epsilon_1}(2a + \epsilon_+ - m)
         + \gamma_{2\epsilon_1, \epsilon_2-\epsilon_1}(- 2a + \epsilon_+ - m)
         + (\epsilon_1 \leftrightarrow \epsilon_2) \right],
           \label{pertvec}
    \eea
  where the definition of the gamma function is presented in Appendix \ref{sec:identity}.
  These results are obtained by adapting the approach of \cite{GL} to our case.
  Since $k_1$ can be integer or half-integer, the instanton partition function can also be divided into 
  an ``even" sector with integer $k_1$ and an ``odd" sector with half-integer $k_1$.
  Since matter in the (anti-)fundamental representation couples to the $\mathcal{O}(k_\alpha C)$ line bundles, 
  its perturbative contribution is different in the two different sectors, giving
    \bea
    Z^{{\rm fund}, even}_{pert}(a, m)
    &=&    \prod_{\alpha=1,2} \exp \left[\gamma_{2\epsilon_1, \epsilon_2-\epsilon_1}(a_{\alpha} + \epsilon_+ - m)
         + (\epsilon_1 \leftrightarrow \epsilon_2) \right],
           \nonumber \\
    Z^{{\rm anti-fund}, even}_{pert}(a,m)
    &=&    \prod_{\alpha=1,2} \exp \left[\gamma_{2\epsilon_1, \epsilon_2-\epsilon_1}(a_{\alpha} + m)
         + (\epsilon_1 \leftrightarrow \epsilon_2) \right]
           \label{pertfund}
    \eea
  with $\vec{a}=(a,-a)$,
  for the even sector, and
    \bea
    Z^{{\rm fund}, odd}_{pert}(a, m)
    &=&    \prod_{\alpha=1,2} \exp \left[\gamma_{2\epsilon_1, \epsilon_2-\epsilon_1}(a_{\alpha} + \epsilon_+ - m + \epsilon_1)
         + (\epsilon_1 \leftrightarrow \epsilon_2) \right],
           \nonumber \\
    Z^{{\rm anti-fund}, odd}_{pert}(a,m)
    &=&    \prod_{\alpha=1,2} \exp \left[\gamma_{2\epsilon_1, \epsilon_2-\epsilon_1}(a_{\alpha} + m + \epsilon_1)
         + (\epsilon_1 \leftrightarrow \epsilon_2) \right]
           \label{pertfundodd}
    \eea
  for the odd sector.
  The contribution of the bifundamental hypermultiplet can also be written as
    \bea
    Z^{{\rm bifund}, even}_{pert}(a, \tilde{a}; m)
    &=&    \prod_{\alpha=1,2} \prod_{\beta' = 1,2} 
           \exp \left[\gamma_{2\epsilon_1, \epsilon_2-\epsilon_1}(a_{\alpha} - \tilde{a}_{\beta'} 
         + \epsilon_+ - m)
         + (\epsilon_1 \leftrightarrow \epsilon_2) \right],
           \label{pertbifundeven} \\
    Z^{{\rm bifund}, odd}_{pert}(a, \tilde{a}; m)
    &=&    \prod_{\alpha=1,2} \prod_{\beta' = 1,2} 
           \exp \left[\gamma_{2\epsilon_1, \epsilon_2-\epsilon_1}(a_{\alpha} - \tilde{a}_{\beta'} 
         + \epsilon_+ - m + \epsilon_1)
         + (\epsilon_1 \leftrightarrow \epsilon_2) \right],
    \eea
  with $\vec{a} = (a, -a)$ and $\vec{\tilde{a}} = (\tilde{a}, - \tilde{a})$.
  Note that the instanton partition function including bifundamental hypermultiplets
  depends on $\vec{k}$ and $\vec{\tilde{k}}$.
  The even (odd) sector of the bifundamental corresponds to the case 
  where $k_{11'} (=k_1 - \tilde{k}_1)$ is (half-)integer, because we fixed $k=\tilde{k}=0$.
  
  By using formul\ae (\ref{f1}), (\ref{f3}) and (\ref{f2}), it is straightforward to see that
    \bea
    & &
    \exp \left[- \gamma_{2\epsilon_1, \epsilon_2-\epsilon_1}(2a - 2k_{12} \epsilon_1)
    - \gamma_{2\epsilon_1, \epsilon_2-\epsilon_1}(2a + \epsilon_+ - 2k_{12} \epsilon_1) 
    - (\epsilon_1 \leftrightarrow \epsilon_2)\right]
           \nonumber \\
    & &    ~~~~~~~~~~~~~~~~~~~~~
     =     (-\Lambda^2)^{(k_{12})^2} ~ \ell_{{\rm vector}}(a) Z^{{\rm vector}}_{pert}(a, m) 
           \label{f1'}
    \eea
  for the vector part, 
    \bea
    & &
    \exp \left[\gamma_{2\epsilon_1, \epsilon_2-\epsilon_1}(2a + \epsilon_+ - m - 2k_{12} \epsilon_1)
    + \gamma_{2\epsilon_1, \epsilon_2-\epsilon_1}(- 2a + \epsilon_+ - m + 2k_{12} \epsilon_1) 
    + (\epsilon_1 \leftrightarrow \epsilon_2)\right]
           \nonumber \\
    & &    ~~~~~~~~~~~~~~~~~~~~~
     =     \Lambda^{-2(k_{12})^2} ~ \ell_{{\rm adj}}(a, m) Z^{{\rm adj}}_{pert}(a, m) 
    \eea
  for the adjoint, 
    \bea
    & &
    \prod_{\alpha=1,2} \prod_{\beta' = 1,2} 
    \exp \left[\gamma_{2\epsilon_1, \epsilon_2-\epsilon_1}(a_\alpha - \tilde{a}_{\beta'}
    + \epsilon_+ - m - 2k_{\alpha \beta'} \epsilon_1)
    + (\epsilon_1 \leftrightarrow \epsilon_2)\right]
           \nonumber \\
    & &    ~~~~
     =     \prod_{\alpha=1,2} \prod_{\beta' = 1,2} \left( \Lambda^{-2(k_{\alpha \beta'})^2} \right) ~ 
           \ell_{{\rm bifund}}(a, \tilde{a}; m) \times
           \left\{ \begin{array}{ll}
           \displaystyle{Z^{{\rm bifund}, even}_{pert}(a, \tilde{a}; m) }, &\quad k_{11'} \in \mathbb{Z} \\
           \displaystyle{Z^{{\rm bifund}, odd}_{pert}(a, \tilde{a}; m)}, &\quad k_{11'} \in \mathbb{Z} + \frac{1}{2}   
           \end{array} \right.
           \label{f2''}
    \eea
  for the bifundamental, and
    \bea
    & &
    \prod_{\alpha = 1,2} 
    \exp \left[\gamma_{2\epsilon_1, \epsilon_2-\epsilon_1}(a_{\alpha} + \epsilon_+ - m - 2k_\alpha \epsilon_1)
         + (\epsilon_1 \leftrightarrow \epsilon_2) \right]
           \nonumber \\
    & &    ~~~~~~~~~~~~
     =     \Lambda^{-(k_1)^2 - (k_2)^2} ~ \ell_{{\rm fund}}(a, m) \times
           \left\{ \begin{array}{ll}
           \displaystyle{(-1)^{(k_1)^2 + (k_2)^2} ~Z^{{\rm fund}, even}_{pert}(a, m)}, &\quad k_1 \in \mathbb{Z} \\
           \displaystyle{(-1)^{(k_1)^2 + (k_2)^2 - 1/2} ~Z^{{\rm fund}, odd}_{pert}(a, m)}, &\quad k_1 \in \mathbb{Z} + \frac{1}{2}   
           \end{array} \right.
           \label{f2'}
    \eea
  for the fundamental part.

  For theories with vanishing beta function, we set $\Lambda=1$.
  In this case, we have to multiply further by the classical part of the partition function:
    \bea
    Z_{cl}
     =     \prod_{s=1}^n q_s^{-\frac{(a^s)^2}{2 \epsilon_1 \epsilon_2}}.
           \label{class}
    \eea
  By combining all together, namely the classical, the perturbative and the instanton parts,
  we obtain the blow-up formula
    \bea
    & &    {\mathcal Z}_{full}^{k^s=0} (\epsilon_1,\epsilon_2,\vec{a}^s; q_s)
           \label{blowup} \\
    & &    ~~~
     =     \sum_{ \vec{k}^s | k^s = 0} (-1)^{(k_{12}^s)^2} 
           Z_{full}^{\mathbb{C}^2} \left(2\epsilon_1,\epsilon_2-\epsilon_1, \vec{a}^s - 2\epsilon_1 \vec{k}^s, q_s \right)
           Z_{full}^{\mathbb{C}^2} \left(\epsilon_1-\epsilon_2,2\epsilon_2, \vec{a}^s - 2\epsilon_2 \vec{k}^s, q_s\right),
    \nonumber \eea
  where
    \bea
    Z_{full}^{\mathbb{C}^2} \left(\epsilon_1,\epsilon_2, \vec{a}^s, q_s \right)
     =     Z_{cl}^{\mathbb{C}^2} \left(\epsilon_1,\epsilon_2, \vec{a}^s, q_s \right)
           Z_{pert}^{\mathbb{C}^2}\left(\epsilon_1,\epsilon_2, \vec{a}^s, q_s \right)
           Z_{inst}^{\mathbb{C}^2} \left(\epsilon_1,\epsilon_2, \vec{a}^s, q_s \right).
           \label{zalefull}
    \eea
  We have defined the classical and the perturbative parts of the partition function on $\mathbb{C}^2$ as follows:
    \bea
    Z_{cl}^{\mathbb{C}^2} \left(\epsilon_1,\epsilon_2, \vec{a}^s, q_s \right)
     =     \prod_{s=1}^n q_s^{-\frac{(a^s)^2}{\epsilon_1 \epsilon_2}},
    \eea
  and the perturbative part is the same as the ones in (\ref{pertvec}), (\ref{pertfund}) and (\ref{pertbifundeven}).
  Notice that the factors $q^{\frac{(k_{12}^s)^2}{2}}$ and $\ell$ in the instanton part (\ref{zale2}) were absorbed, 
  respectively, by the inclusions of the classical and the perturbative parts
  with shifted arguments: $\epsilon_1 \rightarrow 2\epsilon_1$ and $\epsilon_2 \rightarrow \epsilon_2 - \epsilon_1$
  and $\epsilon_{1,2}$ exchanged.
  
  In the case of the asymptotically free theory with one $SU(2)$ gauge group, $q = \Lambda^{b_0}$ where $b_0 = 4 - N_f$, 
  and there is no classical part.
  Then, the $\Lambda$ dependence in (\ref{f1'}), (\ref{f2''}) and (\ref{f2'}) leads to 
  $\Lambda^{\frac{N_f}{2} k^2} \Lambda^{- \frac{b_0}{2} k_{12}^2}$.
  The last factor cancels the $q^{\frac{k_{12}^2}{2}}$ in the instanton part (\ref{zale2}).
  Therefore, the final form of the full partition function (\ref{zalefull}) is universal 
  for the theory with $b_0 \geq 0$ up to a sign.
  
  The blow-up formula (\ref{blowup}) is the analog of the one found in \cite{NY-I} 
  in the $\mathcal{O}_{\mathbb{P}^1}(-1)$ case. 
  The existence of blow-up formulae for $\mathcal{N}=2$ gauge theories on toric manifolds 
  was suggested in \cite{Nekr-local}.

\section{Four-point super Liouville correlator on the sphere}
\label{sec:superLiouville}
  We are now ready to compare the gauge theory partition functions on the $\CO_{\mathbb{P}^1}(-2)$ surface 
  with correlators of the super Liouville conformal field theory.
  As we saw in the previous section, the instanton moduli spaces are classified by $k$.
  The (half-)integer classes correspond to flat connections with (non)trivial holonomy at infinity.
  It is then natural that the integer and half-integer classes are on different footings in the super Liouville theory:
  we expect that they correspond to the Neveu-Schwarz (NS) and Ramond (R) sectors respectively.
  In this paper we focus on the former sector and see the relation with the gauge theory.
  In the present section, we consider the four-point correlation function on the sphere 
  and show that this is related to the full partition function of $SU(2)$ gauge theory with four flavors,
  by concentrating on the case with $k=0$, which is the simplest sector to check.
  
  The $\mathcal{N}=1$ superconformal symmetry of Liouville is generated by the holomorphic currents $T(z)$, $G(z)$ 
  and their anti-holomorphic counterparts.
  The algebra is
    \bea
    \left[ L_m, L_n \right]
    &=&    (m - n) L_{m+n} + \frac{\hat{c}}{8} (m^3 - m) \delta_{m+n, 0},
           \nonumber \\
    \left[ L_n, G_k  \right]
    &=&    \left( \frac{n}{2} - k \right) G_{n+k},
           \nonumber \\
    \{ G_k, G_l \}
    &=&    2 L_{k+l} + \frac{\hat{c}}{2} \left( k^2 - \frac{1}{4} \right) \delta_{k+l,0},
    \eea
  and the same for the right part $\bar{L}_n$ and $\bar{G}_k$.
  In our notation the central charge is $\hat{c} = 1 + 2 Q^2$ with $Q = b + 1/b$.
  The NS sector which we focus on corresponds to the case with $k, l$ are half-integers.
  
  The primary field $V_\alpha(x)$ corresponds to the highest weight vector satisfying
    \bea
    L_n V_\alpha
     =     0, ~~~
    G_k V_\alpha
     =     0, ~~ {\rm for}~k,n>0~~~{\rm and}~~~
    L_0 V_\alpha
     =     \Delta V_\alpha,
    \eea
  where the conformal dimension of the primary is
    \bea
    \Delta
     =    \frac{Q^2}{8} - \frac{\alpha^2}{2},
    \eea
  and similar for the anti-holomorphic part.
  The super-Verma module $\mathcal{V}$ is formed by the descendants $V_{KM} = L_{-M} G_{-K} V$ 
  which are obtained by acting with the raising operators 
  $L_{-M} = L_{-m_1} L_{-m_2} L_{-m_3} \cdots$, where $\{m_i\}$ are positive integers with $m_1 < m_2 < \cdots$,
  and $G_{-K} = G_{-k_1}G_{-k_2}G_{-k_3} \cdots$, where $\{k_i\}$ are positive half-integers with $k_1 < k_2 < \cdots$.
  
  The four-point correlation function is
    \bea
    \left< V_1(\infty) V_2(1) V_3(q) V_4(0) \right>
    &=&    \int d P \Bigg[ \mathbb{C}_{\alpha_1, \alpha_2, iP} 
           \mathbb{C}_{\alpha_3, \alpha_4, - iP} 
           \left| q^{\Delta - \Delta_3 - \Delta_4} \CB_{0,4}^{even} (\Delta_i, q) \right|^2
           \nonumber \\
    & &    ~~~~~~~~
         - \tilde{\mathbb{C}}_{\alpha_1, \alpha_2, iP} \tilde{\mathbb{C}}_{\alpha_3, \alpha_4,-iP} 
           \left| q^{\Delta - \Delta_3 - \Delta_4} \CB_{0,4}^{odd} (\Delta_i, q) \right|^2 \Bigg],
           \label{4point}
    \eea
  up to an irrelevant constant.
  The modulus $q$ will be identified with the gauge coupling constant $q = e^{2 \pi i \tau}$.
  The expressions of the three-point functions
    \bea
    \mathbb{C}_{\alpha_1, \alpha_2, \alpha_3}
     =     C_{\alpha_1+Q/2, \alpha_2+Q/2, \alpha_3+Q/2}, 
           ~~~~ 
    \tilde{\mathbb{C}}_{\alpha_1, \alpha_2, \alpha_3}
     =   - \tilde{C}_{\alpha_1+Q/2, \alpha_2+Q/2, \alpha_3+Q/2}
    \eea
  are given in Appendix \ref{sec:superVirasoro}.
  We divide the conformal block $\CB_{0,4}$ into the even and the odd sectors which are expanded in $q$ as
    \bea
    \CB_{0,4}^{even} 
     =     \sum_{n \in \mathbb{N}} B_{n} q^n, ~~~
    \CB_{0,4}^{odd} 
     =     \sum_{n \in \mathbb{N} + \frac{1}{2}} B_{n} q^{n}.
    \eea
  The lower order terms are computed as \cite{HJS, BBNZ, Belavin}
    \bea
    B_{\frac{1}{2}}
    &=&    \frac{1}{2 \Delta},~~~
    B_1
     =     \frac{(\Delta + \Delta_2 - \Delta_1)(\Delta + \Delta_3 - \Delta_4)}{2 \Delta},
           \\
    B_{\frac{3}{2}}
    &=&    \frac{(1 + 2 \Delta + 2 \Delta_2 - 2 \Delta_1) (1 +2 \Delta + 2 \Delta_3 - 2 \Delta_4)}{8 \Delta (1 + 2 \Delta)}
         + \frac{4(\Delta_2 - \Delta_1)(\Delta_3-\Delta_4)}{(\hat{c} + 2 (-3+\hat{c}) \Delta + 4 \Delta^2)(1 + 2 \Delta)},
           \nonumber 
    \eea
  and so on.
  
  Let us first compare the conformal blocks with the Nekrasov instanton partition function
  of the gauge theory with two fundamental fields with masses $\mu_{3,4}$ 
  and two anti-fundamental fields with masses $\mu_{1, 2}$.
  We first redefine the mass parameters to the ones associated with the $SU(2)^4$ flavor symmetries:
    \bea
    & &    \mu_1
     =     n_0 + m_0 + \frac{\epsilon_+}{2},~~~
    \mu_2
     =     n_0 - m_0 + \frac{\epsilon_+}{2}
           \nonumber \\
    & &    \mu_3
     =     n_1 + m_1 + \frac{\epsilon_+}{2},~~~
    \mu_4
     =     n_1 - m_1 + \frac{\epsilon_+}{2}
    \eea
  Then the instanton partition function is written as
    \bea
    \CZ^{ALE}_{inst}
     =     \sum_{k \in \mathbb{Z}} q^{\frac{k^2}{2}} z^k \CZ^{k}_{inst}
    \eea
  where $\CZ^{k}_{inst}$ is given by (\ref{zale2}), whose $\ell$ factor includes the contributions
  for the vector, the two fundamentals and the two anti-fundamentals.
  Note that $\ell$ in this case depends on both the sum $k$ and the difference $k_{12}$.
  We calculate $\CZ^{k}_{inst}$ with $k=0$ here.
  
  The conformal block above agrees with the instanton partition function 
  by the following relation\footnote{We have checked this relation up to terms of order $q^{5/2}$.} 
  \cite{BBB}:
    \bea
    \CZ^{k=0}_{inst}
    &=&    (1 - q)^{(\frac{\epsilon_+}{2} + n_0)(\frac{\epsilon_+}{2} - n_1)} 
           \left(\CB_{0,4}^{even} + \frac{1}{2} \CB_{0,4}^{odd} \right)
           \label{relation4}
    \eea
  under the identification of the parameters:
    \bea
    & &    \epsilon_1 
     =     b, ~~~
    \epsilon_1
     =     1/b, ~~~
    iP
     =     a,~~~
           \nonumber \\
    & &    \alpha_1
     =     m_0, ~~~
    \alpha_2
     =     n_0,~~~
    \alpha_3
     =     n_1, ~~~
    \alpha_4
     =     m_1.
           \label{identification4}
    \eea
  where $\alpha_i$ are the external momenta.
  The factor in front of the conformal block in (\ref{relation4}) is understood as the $U(1)$ factor
  as introduced in \cite{AGT}.
  Indeed, the conformal block is invariant under the transformation corresponding to the Weyl reflection
  of the flavor symmetry: $n_i \rightarrow - n_i$ and $m_i \rightarrow - m_i$ ($i = 0, 1$).
  
  Let us then compare the remaining part of the correlation function (\ref{4point}).
  It is easy to check that 
    \bea
    \mathbb{C}_{\alpha_1, \alpha_2, iP} \mathbb{C}_{\alpha_3, \alpha_4, - iP}
     =     C_0
           \left| Z_{pert}^{{\rm vector}}(a) \prod_{i=1,2} Z_{pert}^{{\rm anti-fund}, even}(a, \mu_i)
           \prod_{i=3,4} Z_{pert}^{{\rm fund}, even} (a, \mu_i) \right|^2
    \eea
  for the even sector, and 
    \bea
    \tilde{\mathbb{C}}_{\alpha_1, \alpha_2, iP} \tilde{\mathbb{C}}_{\alpha_3, \alpha_4, - iP}
     =   - 4 C_0
           \left| Z_{pert}^{{\rm vector}}(a) \prod_{i=1,2} Z_{pert}^{{\rm anti-fund}, odd}(a, \mu_i)
           \prod_{i=3,4} Z_{pert}^{{\rm fund}, odd} (a, \mu_i) \right|^2
           \label{threepoint4}
    \eea
  for the odd sector,
  where $C_0$ is a factor which does not depend on the internal momentum (or the vev $a$).
  Notice that the difference between the even and the odd sectors in CFT is related with 
  the difference between (\ref{pertfund}) and (\ref{pertfundodd}) of the gauge theory, 
  which correspond to the sectors where $k_1$ is integer and half-integer, respectively.
  
  Also, the factor in front of the conformal block in (\ref{4point}) is identified with the classical part
  of the partition function (\ref{class}):
  $q^{\Delta - \Delta_3 - \Delta_4} \sim q^{- \frac{a^2}{2}}$, up to an $a$ independent factor.
  Therefore, by combining all together, we can see that 
  the four-point correlation function of the super Liouville theory is written as 
  the integral over the internal momentum $P$ with integrand
  $\left| \mathcal{Z}_{full}^{k=0}(a; q) \right|^2$.
  Here we have used that the gauge theory partition function is rewritten as (\ref{zalefull}).

\section{Five-point super Liouville correlator on the sphere}
\label{sec:superLiouville5}
  We consider in this section five-point correlation function on the sphere
  and compare it with the full partition function of $SU(2)^2$ quiver gauge theory 
  with two fundamentals in the first $SU(2)$, two anti-fundamentals in the second $SU(2)$ and one bifundamental.
  
  The correlation function is written as\footnote{In this section and in the following 
                                                  we normalize the three-point functions in the odd-odd sector 
                                                  up to factors 
                                                  $(\Delta_1-\Delta_2 + \Delta_3)
                                                  = \frac{<(G_{-1/2}V_1) V_2 (G_{-1/2} V_3)>}{<V_1 V_2 V_3>}$ 
                                                  which get included in the conformal block. 
                                                  This normalization is more natural for the comparison 
                                                  with the gauge theory.}
    \bea
    \left< \prod_{i=1}^5 V_{\alpha_i} (z_i) \right>
    &=&    \int d P_1 dP_2 
           |(q_1 q_2)^{\tilde{\Delta} - \Delta_4 - \Delta_5} q_1^{\Delta - \tilde{\Delta} - \Delta_3} |^2
           \label{5pointcorrelator} \\
    & &    \Bigg[ \mathbb{C}_{\alpha_1, \alpha_2, iP_1} 
           \mathbb{C}_{- iP_2, \alpha_3, iP_2} \mathbb{C}_{- iP_2, \alpha_4, \alpha_5} |\CB^{e,e}_{0,5}|^2
         - \mathbb{C}_{\alpha_1, \alpha_2, iP_1} 
           \tilde{\mathbb{C}}_{- iP_2, \alpha_3, iP_2} \tilde{\mathbb{C}}_{- iP_2, \alpha_4, \alpha_5} |\CB_{0,5}^{e,o}|^2
           \nonumber \\
    & &  - \tilde{\mathbb{C}}_{\alpha_1, \alpha_2, iP_1} 
           \tilde{\mathbb{C}}_{- iP_2, \alpha_3, iP_2} \mathbb{C}_{- iP_2, \alpha_4, \alpha_5} |\CB_{0,5}^{o,e}|^2
         - \tilde{\mathbb{C}}_{\alpha_1, \alpha_2, iP_1} 
           \mathbb{C}_{- iP_2, \alpha_3, iP_2} \tilde{\mathbb{C}}_{- iP_2, \alpha_4, \alpha_5} |\CB_{0,5}^{o,o}|^2
           \bigg],
           \nonumber 
    \eea
  where $P_{1,2}$ are the internal momenta and we have chosen the coordinates of the insertions as
  $z_1 = \infty, z_2 = 1, z_3 = q_1, z_4 = q_1 q_2$ and $z_5 = 0$.
  The moduli of the sphere $q_i$ will be identified with the gauge coupling constants of the gauge theory
  $q_i = e^{2 \pi i \tau_i}$.
  The indices, $e$ and $o$, of the conformal block $\CB_{0,5}$ denote the even and the odd sectors. 
  Namely the first index $e$ (or $o$) means that the conformal block includes (half-)integer powers in $q_1$
  and the second in $q_2$.
  We can calculate their coefficients as
    \bea
    B_{1,0}
    &=&    \frac{(\Delta + \Delta_2 - \Delta_1)(\Delta + \Delta_3 - \tilde{\Delta})}{2 \Delta},
           ~~~
    B_{0,1}
     =     \frac{(\tilde{\Delta} + \Delta_3 - \Delta)(\tilde{\Delta} + \Delta_4 - \Delta_5)}{2 \tilde{\Delta}},
           \nonumber \\
    B_{\frac{1}{2}, 0}
    &=&    \frac{1}{2 \Delta}, 
           ~~~
    B_{\frac{1}{2}, 1}
     =     \frac{(\tilde{\Delta} + \Delta_3 - \Delta - \frac{1}{2})(\tilde{\Delta} + \Delta_4 - \Delta_5)}
           {4 \Delta \tilde{\Delta}},
           \nonumber \\
    B_{0, \frac{1}{2}}
    &=&    \frac{1}{2 \tilde{\Delta}},
           ~~~
    B_{1, \frac{1}{2}}
     =     \frac{(\Delta + \Delta_2 - \Delta_1)(\Delta + \Delta_3 - \tilde{\Delta} - \frac{1}{2})}
           {4 \Delta \tilde{\Delta}},
           ~~~
    B_{\frac{1}{2}, \frac{1}{2}}
     =     \frac{\Delta - \Delta_3 + \tilde{\Delta}}{4 \Delta \tilde{\Delta}},
    \eea
  and so on.
  
  We now first compare these conformal blocks with the instanton partition function of $SU(2)^2$ gauge theory. 
  We denote the masses of the anti-fundamentals, the fundamentals and the bifundamental by
  $\mu_{1,2}$, $\mu_{4,5}$ and $\mu_3$ respectively.
  The vevs of the vector multiplet scalars of two $SU(2)$ are $a$ and $\tilde{a}$.
  The partition function is
    \bea
    \mathcal{Z}_{inst}^{ALE}
     =     \sum_{k \in \mathbb{Z}} \sum_{\tilde{k} \in \mathbb{Z}} q_1^{\frac{k^2}{2}} q_2^{\frac{\tilde{k}^2}{2}}
           z^{k} \tilde{z}^{\tilde{k}} \mathcal{Z}_{inst}^{k, \tilde{k}}.
    \eea
  We propose that the instanton partition function with fixed first Chern classes $k=0$ and $\tilde{k}=0$ is related 
  to the five-point conformal block as\footnote{We have checked this relation up to terms of order $q_1^{a}q_2^b$
                                                with $a+b = 2$.} 
    \bea
    \mathcal{Z}_{inst}^{k=0, \tilde{k} = 0}
     =     Z_{U(1)} \left( \CB^{e,e}_{0,5} + \frac{1}{2} (\CB^{e,o}_{0,5} + \CB^{o,e}_{0,5} + \CB^{o,o}_{0,5}) \right),
    \eea
  where the $U(1)$ factor is
    \bea
    Z_{U(1)}
     =     (1 - q_1)^{(\frac{\epsilon_+}{2} + n_0)( \frac{\epsilon_+}{2} - m_3)} 
           (1 - q_1 q_2)^{(\frac{\epsilon_+}{2} + m_3)( \frac{\epsilon_+}{2} - n_1)} 
           (1 - q_2)^{(\frac{\epsilon_+}{2} + n_0)( \frac{\epsilon_+}{2} - n_1)}.
    \label{iuuan}\eea
  The identification of the parameters are similar to (\ref{identification4}):
    \bea
    iP_1
    &=&    a,
           ~~~~
    iP_2
     =     \tilde{a},
           \nonumber \\
    \alpha_1
    &=&    m_0,
           ~~~
    \alpha_2
     =     n_0,
           ~~~
    \alpha_3
     =     m_3,
           ~~~
    \alpha_4
     =     n_1,
           ~~~
    \alpha_5
     =     m_1.
    \eea
  We have checked this relation in lower orders in $q_1$ and $q_2$.
  
  Next we consider the three-point function.
  We can show that for the even-even sector
    \bea
    & &
    \mathbb{C}_{\alpha_1, \alpha_2, iP_1} \mathbb{C}_{-iP_1, \alpha_3, iP_2} \mathbb{C}_{\alpha_4, \alpha_5, - iP_2}
           \label{threepoint5} \\
    & &    ~~~
     =     \tilde{C}_0
           \left| Z_{pert}^{{\rm vector}}(a) Z_{pert}^{{\rm vector}}(\tilde{a}) 
           \prod_{i=1,2} Z_{pert}^{{\rm anti-fund}, even}(a, \mu_i)
           Z_{pert}^{{\rm bifund}, even}(a, \tilde{a}; \mu_3)
           \prod_{i=4,5} Z_{pert}^{{\rm fund}, even} (a, \mu_i) \right|^2.
           \nonumber 
    \eea
  To obtain the analog of (\ref{threepoint5}) in the other three sectors, 
  one has to keep into account the numerical $(-4)$ factor appearing in (\ref{threepoint4}) 
  to match the correct normalization as dictated by the blow-up formula (\ref{blowup}).
  
  Finally, the $q_{1,2}$ factor in the first line in (\ref{5pointcorrelator}) can be seen 
  as the classical part of the gauge theory 
  $q_1^{\Delta - \Delta_3 - \Delta_4 - \Delta_5} q_2^{\tilde{\Delta} - \Delta_4 - \Delta_5} 
  \sim q_1^{- \frac{a^2}{2}} q_2^{-\frac{\tilde{a}^2}{2}}$.
  Therefore, we conclude that the integrand of the five-point correlation function can be written as 
  $|\mathcal{Z}_{full}^{k=0, \tilde{k}=0} |^2$.
  
  The arguments presented in this section should generalize to the $n$-point functions on the sphere 
  and the corresponding linear quiver gauge theories. 
  The identification among moduli and gauge couplings should proceed 
  along the same lines as in \cite{AGT} as well as the one among momenta and Coulomb or mass parameters. 
  This should apply to the $U(1)$ factors too, but with different shifted exponents as in (\ref{iuuan}).

\section{One-point super Liouville correlator on the torus}
\label{sec:superLiouvilletorus}
  In this subsection we consider the one-point correlation function on a torus.
  This can be written as
    \bea
    \left< V_{\alpha_1} \right>_{torus}
     =     \int dP ~ \mathbb{C}_{i P, \alpha_1, - i P} \left| q^{\Delta - \frac{c}{24}} \CB_{1,1}(q) \right|^2,
           \label{1point}
    \eea
  again up to an irrelevant constant.
 In (\ref{1point}) the one-point conformal block on the torus is denoted by $\CB_{1,1}(q)$ and expands as 
 $\sum_{n\in\mathbb{N}/2} B_n q^n$ with coefficients
    \bea
    B_{\frac{1}{2}}
    &=&    \frac{- \Delta_1  + 2 \Delta}{2 \Delta},~~~
    B_1
     =     \frac{\Delta_1^2 - \Delta_1 + 2 \Delta}{2 \Delta},
           \nonumber \\
    B_{\frac{3}{2}}
    &=&    2 + \frac{- (\hat{c} + 20 \Delta^2 + 4 \Delta \hat{c}) \Delta_1
         + (4 \Delta^2 + 2 \Delta \hat{c} + 10 \Delta + \hat{c}) \Delta_1^2
         - (2 \Delta + \hat{c}) \Delta_1^3}{2 \Delta (4 \Delta^2 + 2 \Delta \hat{c} - 6 \Delta + \hat{c})}
    \eea
  and so on.
  
  We first consider the relation 
  between the conformal block and the gauge theory partition function of $\CN=2^*$ gauge theory.
  We note that differently from to the $N_f = 4$ case, 
  the $\ell$ factor in this case depends only on the difference $k_{12}$.
  Thus, the instanton partition function factorizes as
    \bea
    {\mathcal Z}^{ALE}_{inst}
     =     \left( \sum_{k \in \mathbb{Z}} q^{\frac{k^2}{2}} z^{k} \right) {\mathcal Z}_{inst}^{k}
     =     \vartheta_3(q; z) {\mathcal Z}_{inst}^{k=0},
    \eea
  where ${\mathcal Z}_{inst}^{k=0}$ is calculated from (\ref{zale2}).
  
  We find that this partition function coincides with the conformal block on the torus with one insertion 
  by the following relation\footnote{We have checked this relation up to terms of order $q^{2}$.}
    \bea
    {\mathcal Z}^{ALE}_{inst}(q;z)
     =     \vartheta_3(q; z) \eta(q)^{- 2 + m (Q - m)} \chi(q) \CB_{1,1}(q)
           \label{relation2}
    \eea
  where $\eta$ and $\chi$ are
    \bea
    \eta(q)
     =     \prod_{n > 0} (1 - q^n), ~~~
    \chi(q)
     =     \prod_{n >0} (1 + q^{n-1/2}).
    \eea
  The identification of the parameters is 
    \bea
    iP
     =     a, ~~~
    \alpha_{1}
     =     m - \frac{\epsilon_+}{2}.
    \eea
  Note that since the factors $\vartheta_3$ and $\chi$ include half-integer powers of $q$,
  this non-trivially mix the even and the odd parts of the conformal block to give the instanton partition function.
  
  We note that the prefactor in (\ref{relation2}) can be written in terms of the character of the affine $SU(2)_2$:
    \bea
    \label{sum}
    {\mathcal Z}^{ALE}_{inst}(q;z)
     =     q^{\frac{1}{16}} \left( \chi^{\widehat{SU}(2)_2}_{[2,0]}(q;z) + \chi^{\widehat{SU}(2)_2}_{[0,2]}(q;z) \right) 
           \eta(q)^{-1 - m (m - Q)} \CB_{1,1}(q)
    \eea
  where $[2,0]$ and $[0,2]$ represent the integrable highest weight representations.
  This result nicely agrees with that of \cite{NT}
  where the sum of the central charges of super Virasoro, affine $\widehat{SU}(2)_2$ and 
  free boson was derived from the M-theory considerations.
  
  Let us briefly comment on this result. 
  First of all one can wonder why the $\chi^{\widehat{SU}(2)_2}_{[1,1]}(q;z)$ term does not appear in the sum (\ref{sum}). 
  The point is that we restricted our computation to $k$ integer 
  and this corresponds in the conformal field theory to restrict to the NS sector. 
  We expect the missing character to arise when including the contribution of instantons with $k$ half-integer, 
  which should be related to the R sector.   
  Secondly, the appearance of the character of the $\widehat{SU}(2)_2$ algebra indicates that the fixed point sector of
  the instanton moduli space on the $\mathbb{Z}_2$ orbifold provides also a realization of this algebra. 
  The result we got in equation (\ref{sum}) suggests 
  that the vertex operator of the entire algebra is non-trivially represented in the super Liouville sector only. 
  In other words, in the punctured torus example we do not see conformal blocks of WZW model 
  appearing in the instanton computation, at least when restricting to the $k$ integer case, namely to the NS sector.    
 
  Finally, let us consider the full correlator on the torus (\ref{1point}).
  The three-point function in (\ref{1point}) can be written as
    \bea
    \mathbb{C}_{i P, \alpha_1, -iP} \left| q^{\Delta - \frac{c}{24}} \right|^2
     =     \widetilde{C}_0 \left| q^{- \frac{a^2}{2}} Z_{pert}^{{\rm vector}}(a) Z_{pert}^{{\rm adj}}(a;m) \right|^2,
    \eea
  where again $\tilde{C}_0$ is an irrelevant constant which does not depend on $a$.
  Therefore, the one-point correlation function on the torus leads to the integral over the vev $a$ 
  of $\left| \mathcal{Z}_{full}^{ALE} \right|^2$.

\subsubsection*{$\CN=4$ partition function}
  The $\CN=4$ partition functions on ALE spaces have been computed and considered 
  from different points of view in relation with affine Lie algebras in the past literature \cite{N=4}.
  Let us briefly consider the $m=0$ limit of the $\CN=2^*$ theory, where $\CN=4$ supersymmetry is recovered.
  We can easily compute the full instanton partition function $\CZ_{inst}^{ALE}$ in this limit as
    \bea
    \CZ_{inst}^{ALE}
     =     \left( \sum_{k \in \mathbb{Z}} q^{\frac{k^2}{2}} z^{k} \right)
           \left( \sum_{k_{12} \in \mathbb{Z}} q^{\frac{k_{12}^2}{2}} \right)
           \eta(q)^{-4}
     =     \vartheta_3(q;z) \eta(q)^{-3} \chi(q)^2
           \label{N=4partition}
    \eea
  where we have used that the instanton partition function on $\mathbb{C}^2$ 
  reduces in this limit to $\eta(q)^{-2}$ and that the $\ell$ factor in (\ref{zale2}) equals $1$ in the massless limit,
  so the sum over $k_{12}$ gets factorized.
  In the last equality, we have used the fact that $\vartheta_3 (q; z=0) = \eta(q) \chi(q)^2$.
  The expression (\ref{N=4partition}) was obtained in \cite{FMP,FMPblack,szabo}.
  As argued in \cite{BBB}, the character of the super Virasoro is
    \bea
    \chi_{super Virasoro} (q)
     =     \eta(q)^{- 1} \chi(q)
    \eea
  which gives a partial check of our result (\ref{relation2}).

\section{Conclusions}
\label{sec:conclusion}
  In this paper we presented a relation between gauge theories on the ALE space $\mathcal{O}_{\mathbb{P}^1}(-2)$ 
  and $\mathcal{N}=1$ super Liouville conformal field theory. 
  We found a blow up formula relating the full partition function of the gauge theory on the ALE space 
  to a convolution of partition functions on the flat space, see Eq.(\ref{blowup}). 
  This simple formula begs for an interpretation in two-dimensional superconformal field theory, 
  possibly relating its correlation functions to the ones of bosonic Liouville theory.
  Moreover, Eq.(\ref{blowup}) could be used to get information on the existence of special geometry relations 
  for gauge theories on the ALE space by using similar arguments to \cite{NY-L}.
  We remark that in this paper we considered the correspondence between the integer $k$ sector of the gauge theory and the NS sector in the CFT. We expect that the half integer $k$ sector will be related to the R sector. Further investigations are needed in this direction.

  On the other side, we found that the super Liouville correlators can be written 
  in terms of four-dimensional gauge theory building blocks.
  Our results then point to a direct interpretation of the correlation functions in super Liouville theory,
  analogous to the one pointed out in \cite{AGT}, 
  as partition functions of the corresponding gauge theories on $S^4/\mathbb{Z}_2$. 
  It would be interesting to pursue this direction more precisely on the gauge theory side 
  along the lines of \cite{Pestun}, in particular carefully analyzing 
  the peculiarities of the gluing conditions in the orbifold case.
  
  We checked the relations we are proposing at lower orders in the expansion in the instanton counting parameters.
  It would be clearly desirable to have an exact, that is to all orders, proof of them 
  by using recursion relations in the super Virasoro conformal block structure \cite{HJS, BBNZ, HJS3, HJS4}, 
  possibly generalizing the approach of \cite{FL,HJS2}. 
  
  Our results open the way to generalize AGT correspondence to more general ADE quotients of $\C^2$
  by comparing with para Liouville/Toda conformal field theories as suggested in \cite{NT}.
  
  It would be very useful in our opinion to find a topological string theory engineering 
  of gauge theories on ALE spaces, 
  possibly implementing a $\Gamma$-equivariant refined BPS counting of M-theory states 
  {\it \`a la} Gopakumar-Vafa \cite{Gopakumar, Hollowood, Iqbal}. 
  Further investigations are also needed on the geometry of M-theory compactifications 
  for this class of theories and its relation to (quantum) Hitchin systems \cite{BT,Teschner,AT,MT,MMM,BMT1}.

Finally, we observe that relations analogous to the AGT correspondence was discussed in
 \cite{Dimofte1,Tera1,Galakhov,Benve,Nishi,Gulotta,Tera2} 
connecting three-dimensional superconformal field theories on (squashed) $S^3$ and 
$SL(2,\mathbb{C})$ Chern-Simons theory on circle bundles over a Riemann surface.
It would be interesting to generalize this correspondence to orbifold spaces.

\section*{Acknowledgments}
We thank H.~Irie, Y.~Tachikawa and M.~Taki for useful discussions and
the participants to the 4th Workshop on Geometric Methods in Theoretical Physics at SISSA 
for stimulating comments.
This research was partly supported by the INFN Research Projects PI14, ``Nonperturbative dynamics of gauge theory" 
and TV12, and by PRIN ``Geometria delle variet\`a algebriche"

\appendix

\section*{Appendix}

\section{$\gamma$ identities}
\label{sec:identity}
  Let us introduce the function
    \beq
    \gamma_{\epsilon_1,\epsilon_2}(x;\Lambda)
     \equiv
           \frac{d}{ds}|_{s=0} \frac{\Lambda^s}{\Gamma(s)}\int_0^\infty \frac{dt}{t}
           t^s \frac{e^{-tx}}{(e^{-\epsilon_1t}-1)(e^{-\epsilon_2t}-1)}
           \label{gamma}
    \eeq
  By changing the argument $\epsilon_1$ and $\epsilon_2$ to the ones in section \ref{sec:gauge}, 
  we have
    \beq
    \gamma_{2\epsilon_1, \epsilon_2-\epsilon_1}(x) + \gamma_{\epsilon_1-\epsilon_2,2\epsilon_2}(x)
    \label{gammaid}
     =     \gamma_{2\epsilon_1, 2\epsilon_2}(x) + \gamma_{2\epsilon_1,2\epsilon_2}(x+\epsilon_1+\epsilon_2),
    \eeq
  where we have suppressed the $\Lambda$ dependence.
  Indeed, from (\ref{gamma}) we get that the l.h.s. of the above expression is
    \bea
    & &    \frac{d}{ds}|_{s=0} \frac{\Lambda^s}{\Gamma(s)}\int_0^\infty \frac{dt}{t}
           t^s {e^{-tx}}\left[\frac{1}{(e^{-2t\epsilon_1}-1)(e^{-t(\epsilon_2-\epsilon_1)}-1)} + 
           \frac{1}{(e^{-2t\epsilon_2}-1)(e^{-t(\epsilon_1-\epsilon_2)}-1)}\right]
           \nonumber\\
    & &    
     =     \frac{d}{ds}|_{s=0} \frac{\Lambda^s}{\Gamma(s)}\int_0^\infty \frac{dt}{t}
           t^s {e^{-tx}}\frac{1+e^{-t(\epsilon_1+\epsilon_2)}}{(e^{-2t\epsilon_1}-1)(e^{-2t\epsilon_2}-1)}  
           \label{gamma2}
    \eea
  from which (\ref{gammaid}) follows.
  One can generalize this identity to the following ones:
  let $s$ be an integer. For $s= \pm 1$, we obtain
    \bea
    \gamma_{2\epsilon_1, \epsilon_2-\epsilon_1}(x + s \epsilon_1)
     + \gamma_{\epsilon_1-\epsilon_2,2\epsilon_2}(x+ s \epsilon_2)
     =      \gamma_{2\epsilon_1, 2\epsilon_2}(x + \epsilon_1)
          + \gamma_{2\epsilon_1,2\epsilon_2}(x + \epsilon_2),
    \eea
  for $s > 1$
    \bea
    \gamma_{2\epsilon_1, \epsilon_2-\epsilon_1}(x + s \epsilon_1)
     + \gamma_{\epsilon_1-\epsilon_2,2\epsilon_2}(x+ s \epsilon_2)
     =      \left( \sum_{{\scriptstyle i,j \geq 0} \atop {\scriptstyle i+j=s}} 
          - \sum_{{\scriptstyle i,j \geq 2} \atop {\scriptstyle i+j=s+2}} \right)
            \gamma_{2\epsilon_1, 2\epsilon_2}(x + i \epsilon_1 + j \epsilon_2),
    \eea
  and for $s < -1$
    \bea
    \gamma_{2\epsilon_1, \epsilon_2-\epsilon_1}(x + s \epsilon_1)
     + \gamma_{\epsilon_1-\epsilon_2,2\epsilon_2}(x+ s \epsilon_2)
     =      \left( \sum_{{\scriptstyle i,j \geq -1} \atop {\scriptstyle i+j=-s-2}} 
          - \sum_{{\scriptstyle i,j \geq 1} \atop {\scriptstyle i+j=-s}} \right)
            \gamma_{2\epsilon_1, 2\epsilon_2}(x - i \epsilon_1 - j \epsilon_2),
    \eea
  By using these identities and generalizing the argument in the Appendix in \cite{NY-L}, 
  we can obtain the following useful formula:
    \bea
    & &
    \exp \left[- \gamma_{2\epsilon_1, \epsilon_2-\epsilon_1}
    (a_{\alpha} - \tilde{a}_{\beta'} + \epsilon_+ - m - 2 k_{\alpha \beta'} \epsilon_1)
    - \gamma_{\epsilon_1-\epsilon_2,2\epsilon_2}
    (a_{\alpha} - \tilde{a}_{\beta'} + \epsilon_+ - m - 2 k_{\alpha \beta'} \epsilon_2) \right]
           \label{f1}\\
    & &
     =     \Lambda^{2 (k_{\alpha \beta'})^2} \ell_{\alpha \beta'}
           (a_{\alpha}, k_{\alpha}; \tilde{a}_{\beta'}, \tilde{k}_{\beta'}; m)^{-1}
           \nonumber \\
    & &    \times
           \left\{ \begin{array}{ll}
           \displaystyle{\exp \left[ - \gamma_{2\epsilon_1, \epsilon_2-\epsilon_1}
           (a_{\alpha} - \tilde{a}_{\beta'} + \epsilon_+ - m)
              - \gamma_{\epsilon_1-\epsilon_2,2\epsilon_2}(a_{\alpha} - \tilde{a}_{\beta'} + \epsilon_+ - m) \right]} , 
           &\quad k_{\alpha \beta} \in \mathbb{Z} \\
           \displaystyle{\exp \left[ - \gamma_{2\epsilon_1, \epsilon_2-\epsilon_1}
           (a_{\alpha} - \tilde{a}_{\beta'} + \epsilon_+ - m + \epsilon_1)
           - \gamma_{\epsilon_1-\epsilon_2,2\epsilon_2}(a_{\alpha} - \tilde{a}_{\beta'} + \epsilon_+ - m + \epsilon_2) \right]}, 
           &\quad k_{\alpha \beta} \in \mathbb{Z} + \frac{1}{2}
           \end{array}
           \right.
           \nonumber
    \eea
  and
    \bea
    & &
    \exp \left[- \gamma_{2\epsilon_1, \epsilon_2-\epsilon_1}
    (a_{\alpha} - \tilde{a}_{\beta'} - m - 2 k_{\alpha \beta'} \epsilon_1)
    - \gamma_{\epsilon_1-\epsilon_2,2\epsilon_2}
    (a_{\alpha} - \tilde{a}_{\beta'} - m - 2 k_{\alpha \beta'} \epsilon_2) \right]
           \label{f3}\\
    & &
     =     (-\Lambda^2)^{(k_{\alpha \beta'})^2} 
           \ell_{\beta' \alpha}(\tilde{a}_{\beta'}, \tilde{k}_{\beta'}; a_\alpha, k_\alpha; - m)^{-1} 
           \nonumber \\
    & &    ~~~ \times
           \left\{ \begin{array}{ll}
           \displaystyle{\exp \left[ - \gamma_{2\epsilon_1, \epsilon_2-\epsilon_1}(a_{\alpha} - \tilde{a}_{\beta'} - m)
              - \gamma_{\epsilon_1-\epsilon_2,2\epsilon_2}(a_{\alpha} - \tilde{a}_{\beta'} - m) \right]} , 
           &\quad k_{\alpha \beta} \in \mathbb{Z} \\
           \displaystyle{\exp \left[ - \gamma_{2\epsilon_1, \epsilon_2-\epsilon_1}(a_{\alpha} - \tilde{a}_{\beta'} - m + \epsilon_1)
              - \gamma_{\epsilon_1-\epsilon_2,2\epsilon_2}(a_{\alpha} - \tilde{a}_{\beta'} - m + \epsilon_2) \right]}, 
           &\quad k_{\alpha \beta} \in \mathbb{Z} + \frac{1}{2}
           \end{array}
           \right.
           \nonumber
    \eea
  where $k_{\alpha \beta'} = k_\alpha - \tilde{k}_{\beta'}$, $\ell_{\alpha \beta}$ was defined in (\ref{ellab}) 
  and we have used
    \bea
    \gamma_{2\epsilon_1, 2\epsilon_2}(x + 2 \epsilon_1) + \gamma_{2\epsilon_1,2\epsilon_2}(x + 2\epsilon_2)
    - \gamma_{2\epsilon_1, 2\epsilon_2}(x) + \gamma_{2\epsilon_1,2\epsilon_2}(x + 2 \epsilon_1 + 2 \epsilon_2)
     =     \log \left( \frac{x}{\Lambda} \right),
    \eea
  by which the $\Lambda$ dependent factor appeared.
  Similary, for the (anti-)fundamental matter part, we obtain
    \bea
    & &
    \exp \left[\gamma_{2\epsilon_1, \epsilon_2-\epsilon_1}(x - 2 k_{\alpha} \epsilon_1)
    + \gamma_{\epsilon_1-\epsilon_2,2\epsilon_2}(x -  2 k_{\alpha} \epsilon_2) \right]
           \label{f2}\\
    & &
     =     (-\Lambda^{-1})^{(k_{\alpha})^2} \ell_{\alpha}(x, k_{\alpha}) \times
           \left\{ \begin{array}{ll}
           \displaystyle{\exp \left[ \gamma_{2\epsilon_1, \epsilon_2-\epsilon_1}(x)
           + \gamma_{\epsilon_1-\epsilon_2,2\epsilon_2}(x) \right]} , 
           &\quad k_\alpha \in \mathbb{Z} \\
           \displaystyle{\exp \left[ \gamma_{2\epsilon_1, \epsilon_2-\epsilon_1}(x + \epsilon_1)
           + \gamma_{\epsilon_1-\epsilon_2,2\epsilon_2}(x + \epsilon_2) \right]}, 
           &\quad k_\alpha \in \mathbb{Z} + \frac{1}{2}
           \end{array}
           \right.
           \nonumber
    \eea
  where $\ell_{\alpha}$ was defined in (\ref{ella}).
  In the case of theories with vanishing beta function, $\Lambda$ is set to be $1$.

\section{Three-point functions}
\label{sec:superVirasoro}
  For completeness, we collect the expressions for the three-point functions of super Liouville theory.
  We are interested in the following types of three-point functions
    \bea
    \left< V_{a_1} (z_1) V_{a_2}(z_2) V_{a_3}(z_3) \right>
    &=&    \frac{C_{a_1, a_2, a_3}}{|z_{12}|^{\Delta_{1+2-3}} |z_{23}|^{\Delta_{2+3-1}} |z_{31}|^{\Delta_{3+1-2}}},
           \nonumber \\
    \left< W_{a_1} (z_1) V_{a_2}(z_2) V_{a_3}(z_3) \right>
    &=&    \frac{\tilde{C}_{a_1, a_2, a_3}}
           {|z_{12}|^{\Delta_{1+2-3} + 1/2} |z_{23}|^{\Delta_{2+3-1} - 1/2} |z_{31}|^{\Delta_{3+1-2}+ 1/2}},
    \eea
  where $\Delta_{1+2-3} = \Delta_1 + \Delta_2 - \Delta_3$.
  The expressions for $C$ and $\tilde{C}$ were given in \cite{Poghosian, RS} and also \cite{FH}
    \bea
    C_{a_1, a_2, a_3}
    &=&    A \frac{\Upsilon_{{\rm NS}}(2a_1) \Upsilon_{{\rm NS}}(2a_2) \Upsilon_{{\rm NS}}(2a_3)}
           {\Upsilon_{{\rm NS}}(a-Q) \Upsilon_{{\rm NS}}(a_{1+2-3}) \Upsilon_{{\rm NS}}(a_{2+3-1}) 
           \Upsilon_{{\rm NS}}(a_{3+1-2})},
           \nonumber \\
    \tilde{C}_{a_1, a_2, a_3}
    &=&    2 i A \frac{\Upsilon_{{\rm NS}}(2a_1) \Upsilon_{{\rm NS}}(2a_2) \Upsilon_{{\rm NS}}(2a_3)}
           {\Upsilon_{{\rm R}}(a-Q) \Upsilon_{{\rm R}}(a_{1+2-3}) \Upsilon_{{\rm R}}(a_{2+3-1}) 
           \Upsilon_{{\rm R}}(a_{3+1-2})},
    \eea
  where $a = a_1 + a_2 + a_3$, $\gamma(x) = \Gamma(x)/\Gamma(1-x)$ and
    \bea
    A
    &=&    \left( \pi \mu \gamma \left( \frac{Qb}{2} \right) b^{1 - b^2} \right)^{\frac{Q-a}{b}} \Upsilon_{{\rm NS}}(0)',
           \nonumber \\
    \Upsilon_{{\rm NS}}(x)
    &=&    \Upsilon \left( \frac{x}{2} \right) \Upsilon \left( \frac{x + Q}{2} \right), 
           ~~~~~~
    \Upsilon_{{\rm R}}(x)
     =     \Upsilon \left( \frac{x + b}{2} \right) \Upsilon \left( \frac{x + b^{-1}}{2} \right).
    \eea
  The $\Upsilon$ function can be written in terms of the Barnes' double Gamma function $\Gamma_2$ as
    \bea
    \Upsilon \left(x \right)
     =     \frac{1}{\Gamma_2 (x | b, b^{-1}) \Gamma_2 (Q-x | b, b^{-1})}.
    \eea


\end{document}